\documentclass[12pt]{article}
\usepackage{a4wide}
\usepackage{amssymb}
\usepackage{amsmath}
\usepackage{graphicx}
\usepackage{mathdots}
\usepackage{slashed}
\usepackage{verbatim}
 \def\makeatletter{\catcode`\@=11}% 11:letter
 \makeatletter
 \def\mathbox#1{\hbox{$\m@th#1$}}%
\def\math@ccstyles#1#2#3#4#5#6#7{{\leavevmode
      \setbox0\mathbox{#6#7}%
      \setbox2\mathbox{#4#5}%
      \dimen@ #3%
      \baselineskip\z@\lineskiplimit#1\lineskip\z@
      \vbox{\ialign{##\crcr
             \hfil \kern #2\box2 \hfil\crcr
             \noalign{\kern\dimen@}%
             \hfil\box0\hfil\crcr}}}}
\def\mathaccstyles{\math@ccstyles\maxdimen}
\def\maththroughstyles{\math@ccstyles{-\maxdimen}}
\def\unity%
 {\maththroughstyles{.45\ht0}\z@\displaystyle {\mathchar"006C}\displaystyle 1}
 
\usepackage{array}
\usepackage{amsthm}

 \numberwithin{equation}{section}

\begin{document}

\mbox{}
\vspace{0truecm}
\linespread{1.1}

%%%%%%%%%%%%%%%%%

%\centerline{\Large \bf  Wilson loops in Symmetric Representations}
%
%
%\bigskip
%
%\centerline{\Large \bf of Large Dimensions in ${\cal N}=2$ Supersymmetric Theories }
%
%\bigskip
%
%\centerline{\Large \bf in a new Double-Scaling Limit }
%
%\medskip

\centerline{\Large \bf  Wilson loops in Large Symmetric Representations}

\bigskip

\centerline{\Large \bf through a Double-Scaling Limit }

\medskip

%\centerline{\LARGE \bf } 

\vspace{.4cm}

 \centerline{\LARGE \bf }

\vspace{1.5truecm}

\centerline{
    { \bf D. Rodriguez-Gomez${}^{a,b}$} \footnote{d.rodriguez.gomez@uniovi.es}
   {\bf and}
    { \bf J. G. Russo ${}^{c,d}$} \footnote{jorge.russo@icrea.cat}}

\vspace{1cm}
\centerline{{\it ${}^a$ Department of Physics, Universidad de Oviedo}} \centerline{{\it C/ Federico Garc\'ia Lorca  18, 33007  Oviedo, Spain}}
\medskip
\centerline{{\it ${}^b$  Instituto Universitario de Ciencias y Tecnolog\'ias Espaciales de Asturias (ICTEA)}}\centerline{{\it C/~de la Independencia 13, 33004 Oviedo, Spain.}}
\medskip
\centerline{{\it ${}^c$ Instituci\'o Catalana de Recerca i Estudis Avan\c{c}ats (ICREA)}} \centerline{{\it Pg.~Lluis Companys, 23, 08010 Barcelona, Spain}}
\medskip
\centerline{{\it ${}^d$ Departament de F\' \i sica Cu\' antica i Astrof\'\i sica and Institut de Ci\`encies del Cosmos}} \centerline{{\it Universitat de Barcelona, Mart\'i Franqu\`es, 1, 08028
Barcelona, Spain }}
\vspace{1cm}

\centerline{\bf ABSTRACT}
\medskip 

We derive exact formulas for circular Wilson loops in the $\mathcal{N}=4$  and $\mathcal{N}=2^{* }$ theories with gauge groups $U(N)$ and $SU(N)$ in the $k$-fold symmetrized product representation. The formulas apply in the limit of large $k$ and small Yang-Mills coupling $g$, with fixed effective coupling $\kappa\equiv g^2k$, and for any finite $N$. 
In the $SU(2)$ and $U(2)$ cases,  closed analytic formulas are obtained for any $k$, while the $1/k$ series expansions are asymptotic.
In the $N\gg 1$ limit, with $N\ll k$, there is an overlapping regime where the formulas can be confronted with results from holography. 
Simple formulas for correlation functions between the $k$-symmetric Wilson loops and chiral primary operators are also given.

\noindent 

\newpage

\tableofcontents

\section{Introduction}

Understanding the properties of extended operators in gauge field theories is important as they can encode aspects of the strong coupling dynamics, such as the emergence of confinement or other phases of the theory.
One prominent example is the Wilson line, supported on a line $C$, which is the trace in a certain representation $\mathcal{R}$ of the gauge group of the holonomy of the gauge field along $C$. Wilson line operators can probe  fine details of the theory, including global properties of the gauge group.

Wilson loop operators may as well be regarded as defects in the ambient gauge theory. From this point of view, they define defect quantum field theories and can be studied by standard tools. One such tool is the large charge expansion (see \cite{Gaume:2020bmp} for an introduction and references). 
In the particular case of 4d gauge theories with $\mathcal{N}=2$ supersymmetry, the sector of chiral primary operators (CPO's)
with large $R$-charge $k$ enjoys special simplifications in a double-scaling limit, where $k\to\infty$ and  the Yang-Mills (YM) coupling $g\to 0$, with fixed $g^2k$  \cite{Bourget:2018obm} (see also \cite{Beccaria:2018xxl,Beccaria:2018owt,Beccaria:2020azj}). 
The existence of this limit is not obvious {\it a priori},
since it requires a specific, dominant dependence  $k^L$ for any given loop order $L$ in correlation functions.
It was later shown in  \cite{Grassi:2019txd} that this limit can be viewed as a 't Hooft limit of an auxiliary matrix model.

Some of the large charge techniques have been recently imported to the study of defect QFT's in \cite{Rodriguez-Gomez:2022gbz}, where RG flows on defects in the Wilson-Fisher theory near 4d and 6d have been studied. Morally speaking, the idea is to consider a large number of coincident defects, so that some of the large charge methods can be deployed. This defect may be regarded as an effective description of a large spin impurity \cite{Cuomo:2022xgw,Cuomo:2021kfm}. 

The Wilson loop computes a phase in the partition function induced by the sweep of a 
%extended 
charged particle in a representation ${\cal R}$ along a line $C$. In view of the previous discussion, it is natural to wonder whether Wilson loops in large representations enjoy special simplifications in very much the same spirit. Motivated by this, here we will study circular supersymmetric Wilson loops in large $k$-symmetric representations in $\mathcal{N}=2$   theories with gauge group $U(N)$ and $SU(N)$.
The insertion of these operators admits a description in terms of a defect QFT, as discussed in \cite{Gomis:2006sb,Gomis:2006im,Hoyos:2018jky,Beccaria:2022bcr}.
In particular, in \cite{Beccaria:2022bcr} a $k$-symmetric representation non-supersymmetric Wilson loop was considered in the double-scaling limit of large $k$ and fixed $g^2k$ to study the RG flows in the defect theory. Other interesting aspects of the defect theory
associated with non-supersymmetric Wilson loops are discussed in  \cite{Beccaria:2017rbe,Beccaria:2018ocq,Beccaria:2019dws,Beccaria:2021rmj}. The double-scaling limit was also considered very recently in \cite{Cuomo:2022xgw} to study $k$-symmetric Wilson loops in $SU(2)$ $\mathcal{N}=2$  superconformal QCD.

\section{Wilson loops in the ${\rm Sym}^k(\Box)$ representation \\
and localization}\label{generalities}

We are interested in circular Wilson loops in the ${\rm Sym}^k(\Box)$ representation in $\mathcal{N}=2$ supersymmetric gauge theories in four spacetime dimensions, with unitary ($U(N)$ or $SU(N)$) gauge group. 
In general  $\mathcal{N}=2$  gauge theories,
these loops can be computed through supersymmetric localization \cite{Pestun:2007rz}.
The vacuum expectation value of a Wilson loop in a representation
${\cal R}$, placed on the equator of $\mathbb{S}^4$, is then obtained by
\begin{equation}
    \langle W_{\cal R} \rangle = \langle {\rm Tr}_{\cal R} \, e^{2\pi\phi}\rangle\ ,
\end{equation}
where $\phi ={\rm diag}(a_1,...,a_N)$ parametrizes the Coulomb moduli. For a gauge group $U(N)$, the average is computed by the integral
%\footnote{We choose to normalize with an extra $N^{-1}$ to follow the conventions of \cite{Drukker:2005kx}.} 
 
\begin{equation}
\label{Wkgenericformula}
\langle W_{\cal R}\rangle=\frac{1}{Z_{U(N)}}\,\int d^Na\,\prod_{i<j}(a_i-a_j)^2\,Z_{\rm 1-loop}\,Z_{\rm inst}\,e^{-\frac{8\pi^2}{g^2}\,\sum_{i=1}^N a_i^2}\, W_{\cal R}\,,
\end{equation}
with 
\begin{equation}
Z_{U(N)}=\int d^Na\,\prod_{i<j}(a_i-a_j)^2\,Z_{\rm 1-loop}\,Z_{\rm inst}\,e^{-\frac{8\pi^2}{g^2}\,\sum_{i=1}^N a_i^2}\,.
\end{equation}
When the gauge group is $SU(N)$, one needs to take into account the extra constraint
$\sum_i a_i=0$, as usual.

In the above expressions, $Z_{\rm 1-loop}$ is the one-loop determinant and $Z_{\rm inst}$ is the factor that contains the instanton contributions (it is worth recalling that $Z_{\rm 1-loop}$ and $Z_{\rm inst}$ are symmetric under the permutations of the $a_i$'s). Both factors depend on the specific $\mathcal{N}=2$ theory. In this note we will focus on $\mathcal{N}=4$  and  $\mathcal{N}=2^{* }$ theories (the generalization of our results to any other  $\mathcal{N}=2$ theory with unitary gauge group is straightforward).
In particular, for the ${\mathcal{N}=4}$ theory,

\begin{equation}
    Z_{\rm 1-loop}^{\mathcal{N}=4}=Z_{\rm inst}^{\mathcal{N}=4}=1\, . 
\end{equation}
For the $\mathcal{N}=2^*$ theory, obtained as usual by adding a mass term for the hypermultiplet, one has
\begin{equation}
\label{1loopN2*}
\qquad Z_{\rm 1-loop}^{\mathcal{N}=2^*}=\prod_{i<j}^{N}\frac{H(a_i-a_j)^2}{H(a_i-a_j+M)H(a_i-a_j-M)}\, ,
%\qquad Z_{\rm 1-loop}^{\rn SQCD}=\frac{\prod_{i<j}H(a_i-a_j)^2}{\prod_i H(a_i)^{2N}}\,,
\end{equation}
where
\begin{equation}
    H(x)\equiv e^{-(1+\gamma) x^2}G(1+ix)G(1-ix)=\prod_{n=1}^\infty \left( 1+\frac{x^2}{n^2}\right)^n e^{-\frac{x^2}{n}}\,,
\end{equation}
where $G(x)$ is the Barnes $G$-function. 

In the case of the ${\cal N}=4$ theory,  the partition function reduces to that of the Gaussian
matrix model and one obtains \cite{mehtabook}
\begin{eqnarray}
  \label{zun}
 &&Z_{U(N)}\ =\frac{g^{N^2}}{2^{N^2}(2\pi )^{\frac12N(2N-1)}} G(N+2)\ ,
\\
  \label{zsun}
 &&Z_{SU(N)}=2\sqrt{2\pi N}\frac{g^{N^2-1}}{2^{N^2}(2\pi )^{\frac12N(2N-1)}} G(N+2)\ .
\end{eqnarray}

Lastly, we need to specify the insertion $W_{\mathcal{R}}$ corresponding to the Wilson loop in the desired representation (\textit{e.g.} \cite{Okuyama:2006jc,Hartnoll:2006is,Fiol:2013hna,Chen-Lin:2015dfa,Fiol:2018yuc}), which basically corresponds to its the character. In our case, let us denote by $W_k$ the Wilson loop in the $k$-symmetric representation. To find that, note that the maximal torus of $U(N)$ is $U(1)^N$. Denoting by $z_i$ the fugacity associated to the $i$-th torus, the character of the fundamental representation of $U(N)$ is $\sum_{i=1}^N\,z_i$. The generating function for the symmetrized products is then

\begin{equation}
    F_S(t)={\rm PE}[t\,\sum_{i=1}^Nz_i]=\prod_{i=1}^N\frac{1}{1-t\,z_i}\,,
\end{equation}
where ${\rm PE}$ is the plethystic exponential.\footnote{The plethystic exponential is defined as ${\rm PE}[f(x_1,x_2,\cdots)]=e^{\sum_{k=1}^{\infty}\frac{f(x_1^k,x_2^k,\cdots)}{k}}$.} By definition, the coefficient of $t^k$ in the expansion of $F_S$ is the character of the $k$-fold symmetrized product of fundamental representations. One can extract $W_k$ by using the formula

\begin{equation}
    W_k=\frac{1}{2\pi i}\int \frac{dt}{t^{k+1}}\,F_S(t)\,.
\end{equation}
Computing the integral, one obtains ($z_i=e^{2\pi a_i}$)

\begin{equation}
\label{Wk}
     W_k= \sum_{i=1}^N\,\frac{e^{2\pi (N-1)\,a_i+2k\pi a_i}}{\prod_{j\ne i} \big( e^{2\pi a_i}-e^{2\pi a_j} \big)}\,.
\end{equation}
This formula is equivalent to the expected result (see \textit{e.g.} \cite{Hartnoll:2006is,Chen-Lin:2015dfa})

\begin{equation}
    W_k=\sum_{1\leq i_1\leq i_2\cdots \leq i_k\leq N}e^{2\pi a_{i_1}+2\pi a_{i_2}+\cdots +2\pi a_{i_k}}\,.
\end{equation}
The dimension of the $k$-symmetric representation is 
$$
d_k={\rm dim}\, {\rm S}^k=\frac{(N-1+k)!}{(N-1)!\ k!}\ .
$$
A natural choice of normalization is to define the operator $W_k$ by adding the extra  factor $1/d_k$.
Here we will follow the conventions of \cite{Drukker:2005kx} and
normalize by adding a factor $1/N$. 
Thus, using \eqref{Wk},  the VEV of $W_k$ \eqref{Wkgenericformula} takes the form

\begin{equation}
   \langle W_k\rangle=\frac{1}{N\,Z_N}\,\sum_{i=1}^N \int d^Na\,\prod_{k<l}(a_k-a_l)^2\,Z_{\rm 1-loop}\,Z_{\rm inst}\,e^{-\frac{8\pi^2}{g^2}\sum_{m=1}^N a_m^2}\frac{e^{2\pi (k+N-1)\,a_i}}{\prod_{j\ne i} \big( e^{2\pi a_i}-e^{2\pi a_j}\big)}\,. 
   \nonumber
\end{equation}
By symmetry, the $N$ terms in the sum are equal, therefore we get

\begin{equation}
\label{Wk2}
    \langle W_k\rangle=\frac{1}{Z_{U(N)}} \int d^Na\,\prod_{k<l}(a_k-a_l)^2\,Z_{\rm 1-loop}\,Z_{\rm inst}\,e^{-\frac{8\pi^2}{g^2}\sum_{m=1}^N a_m^2}\frac{e^{2\pi (k+N-1)\,a_N}}{\prod_{j\ne N}\big( e^{2\pi a_N}-e^{2\pi a_j}\big) }\,. 
\end{equation}
When the gauge group is $SU(N)$, the same formula applies upon imposing the constraint $\sum_{i=1}^Na_i=0$ in the integral.

\medskip

Our goal is to study the formula \eqref{Wk2} in the double-scaling limit
\begin{equation}
\label{limitedoble}
  g\to 0\, ,\quad   k\rightarrow \infty \, ,\qquad g^2k=\kappa={\rm fixed}\,.
\end{equation}
We shall see below that in this limit the integral can be computed exactly by  the saddle point method.
An important simplification is that instanton contributions vanish exponentially in this limit, since they are proportional to 
$e^{-\frac{8\pi^2n}{g^2}}= e^{-k\frac{8\pi^2 n}{\kappa}}$.
Since the instanton moduli space does not depend on $k$, there can be no compensating effect to the exponential suppression of the instanton action.
This is the same mechanism as in  the large charge limit of \cite{Bourget:2018obm}. Therefore, upon taking the limit one can set 
$Z_{\rm inst}=1$.

%%%%%%%%%%%%%%%%%%%%%%
\section{$\langle W_k\rangle$ in the ${\cal N}=4$ theory with   $SU(2)$ and $U(2)$}\label{U(2)}

Let us start with the $SU(2)$ case, where $a\equiv a_1=-a_2$. From \eqref{Wk} we get

\begin{equation}
W_k= \frac{e^{2\pi(k+1)a}-e^{-2\pi(k+1)a}}{e^{2\pi a}-e^{-2\pi a}}=\frac{\sinh\big(2\pi(k+1)a\big)}{\sinh(2\pi a)}\,.
\end{equation}
We shall first consider the computation of $\langle W_{k}\rangle$ in the ${\cal N}=4$ theory. As mentioned, in this case $Z_{\rm 1-loop}=1$ and
$Z_{\rm inst}=1$. 
The VEV of the loop in the $k$-symmetric representation is, therefore,

\begin{equation}
    \langle W_{k}\rangle=\frac{1}{2\,Z_{SU(2)}}\,\int da\,4a^2\,e^{-\frac{16\pi^2}{g^2}\,a^2}\,W_{k}\, .
\end{equation}
Substituting the explicit form of $W_k$, we obtain
\begin{equation}
    \langle W_{k}\rangle=\frac{2I_{k+1}}{Z_{SU(2)}}\ \ ,
\end{equation}
 where

\begin{equation}
   I_k \equiv  \int_{-\infty}^\infty da\ \frac{a^2 \sinh(2\pi k a)}{\sinh(2\pi a)}\, e^{-b a^2} \,, \,\qquad b=\frac{16\pi^2}{g^2}\,.
\end{equation}
Remarkably, the integral $I_k$ can be carried out exactly for any integer $k$ in terms of elementary functions. We obtain
\begin{eqnarray}
\label{impares}
   && k=2n+1\ ,\qquad I_{2n+1}=\frac{\sqrt{\pi}}{2b^{\frac52}}\left(b +2\sum_{r=1}^n e^{\frac{4r^2\pi^2}{b}} (b+8r^2\pi^2)\right)\ ,
\\
\label{pares}
    && k=2n+2\ ,\qquad I_{2n+2}=\frac{\sqrt{\pi}}{b^{\frac52}} e^{\frac{\pi^2}{b}}\sum_{r=0}^{n} e^{\frac{4r(r+1)\pi^2}{b}} \left(
b+2(2r+1)^2\pi^2\right)\ ,
\end{eqnarray}
with $n=0,1,2,...$. Then

\begin{eqnarray}
\label{Wimpares}
   &&\langle W_{2n}\rangle=\frac{1}{2} +\sum_{r=1}^n e^{\frac{r^2 g^2}{4}} (1+ \frac{r^2g^2}{2})\ ,
\\
\label{Wpares}
  &&\langle W_{2n+1}\rangle=e^{\frac{g^2}{16}}\sum_{r=0}^{n} e^{\frac{r(r+1)g^2}{4}} \Big(1+(2r+1)^2 \frac{g^2}{8}\Big)\ ,
\end{eqnarray}
where we used $Z_{SU(2)}=g^3/(32\pi^{\frac52})$.
In particular, 

\begin{equation}
\label{nosotros}
\langle W_1\rangle=    e^{\frac{g^2}{16}}\,  (1+\frac{g^2}{8})\ .  
\end{equation}

As a check, $\langle W_1\rangle $ can be compared with the known formula for the circular Wilson loop computed by Drukker and Gross in \cite{Drukker:2000rr}. For $SU(N)$, 
\begin{equation}
    \langle W_1\rangle_{\rm DG} = \frac{2 e^{-\frac{g^2(1+N)}{8N}}}{N! g}
    \int_0^\infty dt e^{-t} t^{N-\frac12} I_1(\sqrt{t} g) =
\frac{e^{-\frac{g^2(N+1)}{8N}}}{N }L^1_{N-1}(-g^2/4)\ .
\end{equation}
For $N=2$, we get
\begin{equation}
    \langle W_1\rangle_{\rm DG}  = \frac{ e^{-\frac{3g^2}{16}}}{ g}
    \int_0^\infty dt e^{-t} t^{\frac32} I_1(\sqrt{t} g) =
\frac{e^{\frac{g^2}{16}}}{2 }L^1_{1}(-g^2/4)=e^{\frac{g^2}{16}}(1+\frac{g^2}{8})
\end{equation}
in agreement with our result \eqref{nosotros}.

\medskip

One can compute correlation functions of Wilson loops (see  \cite{Galvagno:2021bbj} for other examples). Owing to the identity

\begin{equation}
    \frac{\sinh(2\pi (n+1)a)}{\sinh(2\pi a)}\,\frac{\sinh(2\pi(m+1)a)}{\sinh(2\pi a)}=\sum_{k=\frac{|n-m|}{2}}^{\frac{n+m}{2}} \frac{\sinh\big( 2\pi(2k+1)a\big)}{\sinh(2\pi a)}\,,
\end{equation}
it follows that  correlation functions between two Wilson loops satisfy the general relation

\begin{equation}
\label{Wcorrelator}
    \langle W_n\,W_m\rangle=\sum_{k=\frac{|n-m|}{2}}^{\frac{n+m}{2}}\langle W_{2k}\rangle\,,
\end{equation}
where we formally denote $\langle W_0\rangle =1$.

\subsection{The large $k$ limit}

We now consider the double-scaling limit \eqref{limitedoble}.
In this large $k$ limit, the integral can be computed by the saddle-point method.
We have
\begin{equation}
    \langle W_{k}\rangle_{SU(2)}= \frac{2}{ Z_{SU(2)}}\int_{-\infty}^\infty da\ \frac{a^2 }{\sinh(2\pi a)}\, e^{-\frac{16\pi^2}{g^2} a^2+2\pi a(k+1)}\ .
\end{equation}
There is a saddle point at
\begin{equation}
    a_{* }=\frac{g^2(k+1)}{16\pi}\,.
\end{equation}
Using
\begin{equation}
    Z_{SU(2)}=
%4\,\int da\,a^2\,e^{-\frac{16\pi^2}{g^2}\,a^2}=
\frac{\kappa^{\frac{3}{2}}}{32\, \pi^{\frac{5}{2}}\,k^{\frac{3}{2}}}\,,
\end{equation}
we find 
\begin{equation}
\label{sudos}
   \log \langle W_{k}\rangle_{SU(2)}=\frac{k\,\kappa}{16}+\log \frac{k\,\kappa}{16} +\frac{\kappa}{8} 
 -  \log\left( \sinh\frac{\kappa}{8}\right)+O\left(k^{-1}\right)\, .
\end{equation}
\bigskip

It is worth pointing out some salient aspects of the expansion in powers of $1/k$.
Introduce a new integration variable, $x=a-a_*$, 
\begin{equation}
   I_k= e^{\frac{k\kappa}{16}} \int_{-\infty}^\infty dx\ \frac{(x+a_*)^2}{\sinh(2\pi (x+a_*))}\, e^{-\frac{16\pi^2 k}{\kappa} x^2}\, .
\end{equation}
The $1/k$ series is generated by expanding the factor multiplying the exponential in powers of $x$.
With the change of variable, $x^2=t$, one may put this integral in the familiar form used in the Borel analysis. 
The convergence properties of the $1/k$ expansion can be deduced by
 studying the singularities of the integrand. The integrand has poles in the complex plane at
\begin{equation}
    x=-a_* + \frac{i n}{2}
 \ ,\qquad n=\pm 1 , \pm 2, ...
\end{equation}
This implies that the series expansion of $1/{\sinh\big(2\pi (x+a_*)\big)}$ around $x=0$ has a finite radius of convergence, given by
$r_0=\sqrt{a_*^2+1/4}$, corresponding to the value of $|x|$ where the integrand has the first poles at $n=\pm 1$.
The integral over $x$ in each term of the series is of the form $\int dx\,  x^{n-1} e^{-b x^2}$ and gives an extra $n!$ for the $n$-th term. This
proves that the $1/k$ series is asymptotic.

It is interesting to contrast the asymptotic series representation with the compact form given by the exact integration given in (\ref{Wimpares}), (\ref{Wpares}). The compact form, though it involves a finite number of terms, 
is not in the form of a $1/k$ series, because the summation limit involves $k$ itself.

In resurgence theory, asymptotic series may indicate missing non-perturbative contributions.
In  the present case of the ${\cal N}=4$ theory, instanton sectors are not responsible of the asymptotic behavior of the perturbation series  because in this case there are no instantons. The semiclassical field configurations  contributing to  discontinuities across the
Stokes lines are in correspondence with semiclassical solutions for the constant part of the scalar field of the vector multiplet in the one-loop effective action \cite{Aniceto:2014hoa}.

In the ${\cal N}=2^*$ theory, there are instanton contributions of order $e^{-8\pi^2 |n|/g^2}=e^{-8\pi^2 |n|k/\kappa}$.
On the other hand, the $1/k$ series is now different, and much more complicated, since the integrand gets corrected by  the 1-loop determinant factor, 
which itself leads to new singularities in the Borel plane.
It would be interesting to understand how the resurgence analysis works in this case, and the interplay between instanton contributions and the new singularities, in particular, whether  instanton contributions may resurge by a proper treatment of the $1/k$ expansion.

%%%%%%%%%%%%%%%%%%%%%%%%%%%%%%
\subsection{The $U(2)$ case}

When the gauge group is $U(2)$, the relevant VEV of the Wilson loop operator in the $k$-symmetric representation is obtained by
setting  $N=2$ in \eqref{Wk2}. 
This gives
\begin{equation}
     \langle W_k\rangle =\frac{J_k}{2 Z_{U(2)}}\ ,
\end{equation}
where 

\begin{equation}
\label{jk2}
     J_k =2 \int da_1da_2\,(a_1-a_2)^2\,e^{-\frac{8\pi^2}{g^2}(a_1^2+a_2^2)}\frac{e^{2\pi (k+1)a_2}}{e^{2\pi a_2}-e^{2\pi a_1}}\,. 
\end{equation}
By introducing new integration variables, $a_1=x-y,\ a_2=x+y$, $J_k$  takes the form 
\begin{equation}
\label{jotakk}
   J_k= 8\int dy \,y^2\, \frac{e^{-2\pi (k+1) y}}{\sinh(2\pi y) }\, e^{-\frac{16\pi^2}{g^2}\, y^2}\int dx\ e^{-\frac{16\pi^2}{g^2}\, x^2}e^{2\pi k x}\, .
\end{equation}
Computing the Gaussian integral in $x$, we obtain
\begin{equation}
    J_k= I_{k+1} \, \frac{2\,g e^{\frac{g^2 k^2}{16}}}{\sqrt{\pi }}\ .
\end{equation}
Using that 

\begin{equation}
Z_{U(2)}=Z_{SU(2)}\, \frac{g}{2\sqrt{\pi}}\,,
\end{equation}
we arrive at 

\begin{equation}
\label{SU2vsU2}
    \langle W_k\rangle_{U(2)} = e^{\frac{g^2 k^2}{16}}\, \langle W_k\rangle_{SU(2)} \,.
\end{equation}
Substituting into this formula the expression (\ref{sudos}) for $\langle W_k\rangle_{SU(2)}$, 
we obtain the large $k$, fixed $\kappa$ asymptotics, which reads (modulo $1/k$ corrections)

\begin{equation}
    \langle W_{k}\rangle_{U(2)}=  \frac{k\,\kappa}{16}\frac{e^{\frac{k\,\kappa}{8}} e^{\frac{\kappa}{16}} }{\sinh\frac{\kappa}{8}}\, .
\end{equation}
One can reproduce the same result from a saddle-point evaluation of (\ref{jotakk}).

%%%%%%%%%%%%%%%%%%%%%%%%%%%%%%%%%%%%%%%%%

%%%%%%%%%%%%%%%%%%%%%%%%%%%
\section{$\langle W_k\rangle$ in the ${\cal N}=4$ theory with $U(N)$ and $SU(N)$}\label{U(N)}

\subsection{$\langle W_k\rangle$ in the $U(N)$ gauge theory}

It is convenient to write the formula (\ref{Wk2}) for the VEV of the loop   as follows:

\begin{equation}
    \langle W_k\rangle=\frac{e^{\frac{k\kappa}{8}\,(1+\frac{N-1}{k})^2}}{Z_{U(N)}}\,\, \int d^{N}a\,\prod_{k<l}(a_k-a_l)^2\,e^{-k\frac{8\pi^2}{\kappa}\sum_{i=1}^{N-1} a_i^2}\,\left(\frac{e^{-k\frac{8\pi^2}{\kappa}\big(a_N- a_N^{* }\big)^2}}{\prod_{j\ne N}\big( e^{2\pi a_N}-e^{2\pi a_j}\big) }\right)\, ,
\end{equation}
where $\kappa=g^2\,k$  and 
\begin{equation}
    a_N^{* }\equiv \frac{\kappa}{8\pi}(1+\frac{N-1}{k}) \,.
\end{equation}
Let us now take  the double-scaling limit \eqref{limitedoble}
involving $g\to 0$ and $k\to\infty$.
There is a saddle point for $a_N$  at 
\begin{equation}
    a_N^{* }= \frac{\kappa}{8\pi} \,.
\end{equation}
In order to compute the integrals over $a_i$, with $i=1,...,N-1$, it is convenient to 
 introduce new coordinates $x_i=a_i/g$, and expand the integrand in powers of $g$. Because only even powers of $x_i$ survive the integration,
 the next to leading contribution is of order ${\cal O}(g^2)={\cal O}(1/k)$ and
 can be ignored in the limit $k\to \infty$. In this limit, $  \langle W_k\rangle$ is
 exactly determined by the leading term, given by
 
\begin{equation}
    \langle W_k\rangle=\frac{e^{\frac{k\kappa}{8}\,(1+\frac{N-1}{k})^2} }{Z_{U(N)}}\int d^{N-1}a\,\prod_{k<l<N}(a_k-a_l)^2\,e^{-k\frac{8\pi}{\kappa}\sum_{i=1}^{N-1} a_i^2} \int da_N\, a_N^{2(N-1)}\frac{e^{-k\frac{8\pi^2}{\kappa}(a_N-\frac{\kappa}{8\pi})^2}}{(e^{2\pi a_N}-1)^{N-1}}\, , 
\end{equation}
where we have restored the $a_i$ variables.
 Note that, in the Vandermonde determinant, we have replaced $|a_N-a_i|^2$ by $|a_N|^2$, since the difference is an ${\cal O}(1/k)$ contribution.
This approximation assumes finite $N$.
In the infinite $N$, 't Hooft limit, the Vandermonde determinant provides a repulsion between the eigenvalues and the scale for $a_i$ is of order $\sqrt{\lambda}=g\sqrt{N}$, which is finite in the 't Hooft limit.
We shall discuss the case of large $N$ below.

We have kept a term of order $\frac{N-1}{k}$ in the exponential factor outside the integral. This is because it is multiplied by $k$;
it will lead to a finite contribution in the final formula for $\log \langle W_k\rangle$.

The integral over the $a_i$'s factorizes from the integral over $a_N$, giving  a factor $Z_{U(N-1)}$, so we get
\begin{equation}
\label{N=4}
    \langle W_k\rangle=\frac{Z_{U(N-1)}}{Z_{U(N)}}\,e^{\frac{k\kappa}{8}\,(1+\frac{N-1}{k})^2}\, \int da_N\,\Big(\frac{a_N^2}{e^{2\pi a_N}-1}\Big)^{N-1}\,e^{-k\frac{8\pi^2}{\kappa}(a_N-\frac{\kappa}{8\pi})^2}\,. 
\end{equation}
Using (\ref{zun}) and computing the integral by saddle point, we  obtain the following formula for the large $k$ asymptotics of  $\langle W_k\rangle $:

\begin{equation}
\label{result2}
    \langle W_k\rangle =\frac{1}{N!}\, \Big(\frac{k\,\kappa}{4}\Big)^{N-1}\,e^{\frac{k\kappa}{8}\,(1+\frac{N-1}{k})^2}\,e^{-\frac{\kappa\,(N-1)}{4}}\, \Big(1-e^{-\frac{\kappa}{4}}\Big)^{1-N}\, ,
\end{equation}
where we used the fundamental property of the Barnes $G$-function, $N! G(N+1)=G(N+2) $. 
Thus the first terms in the large $k$ expansion for $\log  \langle W_k\rangle$ are

\begin{equation}
\label{logresult}
    \log \langle W_k\rangle=\frac{k\kappa}{8} +(N-1)\log \frac{k\,\kappa}{4} -\log {N!}-(N-1)\log \Big(1-e^{-\frac{\kappa}{4}}\Big)
+{\cal O}(k^{-1})\,.
\end{equation}
where terms ${\cal O}(k^{-1})$ also stands
for terms of order $(N-1)/k$.

\medskip

\subsubsection*{The multiply wound loop}

It is of interest to compare $\langle W_k\rangle $ with the VEV of the $k$-wound circular Wilson loop in the fundamental representation. This corresponds to the insertion of 

\begin{equation}
   {W}_k^{\rm F}=\sum e^{k\,2\pi a_i}\, .
\end{equation}
instead of $W_k$. It is clear that upon defining $a'_i=ka_i$, the computation is just analogous to that of the circular Wilson loop  upon replacing $g$ by $gk$ \cite{Drukker:2000rr}. Thus

\begin{equation}
    \langle {W}_k^{\rm F}\rangle=\frac{1}{N}\,L_{N-1}^1(-\frac{k\kappa}{4})\,e^{\frac{k\kappa}{8}}\,,
\end{equation}
where we have conveniently re-written the result in terms of $\kappa=g^2k$. Let us now consider the large $k$ limit at fixed $\kappa$ and $N$ (we stress that this is a different regime than the one studied in \cite{Drukker:2005kx}).
We obtain

\begin{eqnarray}
\label{multiplywoundresult}
    \langle {W}_k^{\rm F}\rangle&=&\frac{1}{N!}\, \Big(\frac{k\,\kappa}{4}\Big)^{N-1}\,e^{\frac{k\kappa}{8}\,(1+\frac{N-1}{k})^2}\,e^{-\frac{\kappa\,(N-1)}{4}}\Big[ 1-\frac{(N-1)\,\big((N-1)\kappa^2-32N\big)}{8k\kappa}+\cdots\Big]\nonumber \\ & = & \frac{1}{N!}\, \Big(\frac{k\,\kappa}{4}\Big)^{N-1}\,e^{\frac{k\kappa}{8}\,(1+\frac{N-1}{k})^2}\,e^{-\frac{\kappa\,(N-1)}{4}}\left(1+ {\cal O}(k^{-1})\right)\,.
\end{eqnarray}
This is to be compared with \eqref{result2}. Therefore, modulo corrections that vanish in the infinite $k$ limit, one finds the relation
\begin{equation}
\label{kwoundksymm}
  \langle {W}_k\rangle= \Big(1-e^{-\frac{\kappa}{4}}\Big)^{1-N}\, \langle {W}_k^{\rm F}\rangle\ ,\qquad k\gg 1\ .
\end{equation}
In a small $\kappa $ expansion, one has $\langle {W}_k^{\rm F}\rangle \approx\left(\frac{\kappa}{4}\right)^{N-1} \langle {W}_k\rangle$. In a large $\kappa$ expansion, $\langle {W}_k^{\rm F}\rangle $ and $ \langle {W}_k\rangle$ differ
in an infinite series of exponentially small terms $e^{-n\kappa/4}$. We shall return to the interpretation of these exponential terms below.

\subsection{Comparing with holography}

The formula (\ref{logresult}) for $\log \langle W_k\rangle $ can be extended to  $N\gg 1 $ provided  that $k\gg N$. Define $S_k$ by
\begin{equation}
\label{result21}
    \langle W_k\rangle= \frac{e^{-S_k}}{\sqrt{2\pi N}}\, .
\end{equation}
Using Stirling's approximation in (\ref{logresult}), one obtains
\begin{equation}
S_k= -\frac{k\kappa}{8}-N \log\Big(\frac{k\,\kappa}{4N}\Big)-N+N\,\log\Big(1-e^{-\frac{\kappa}{4}}\Big)+\mathcal{O}\left(k^{-1}\right)\,.
\end{equation}
In terms of the standard  't Hooft coupling
$\lambda\equiv g^2N$,

\begin{equation}
    \kappa=\lambda\,\frac{k}{N}\,.
\end{equation}
We will assume large $k$  and large $N$ with fixed and very small $\frac{N}{k}$. In this case the supergravity regime   $\lambda\gg 1$ implies  $\kappa\gg 1$ and we can neglect exponentially suppressed terms ${\cal O}(e^{-\kappa/4})$, so $\langle {W}_k\rangle\to \langle {W}_k^{\rm F}\rangle $. Therefore,  in  the supergravity regime we get

\begin{equation}
\label{result3}
    \langle W_k\rangle= \frac{e^{-S_k}}{\sqrt{2\pi N}}\,,\qquad S_k= -\frac{k\kappa}{8}-N \log\Big(\frac{k\,\kappa}{4N}\Big)-N+\mathcal{O}\left(k^{-1}\right)+{\cal O}(e^{-\kappa/4})\,.
\end{equation}
This result may be compared against the holographic computation in \cite{Drukker:2005kx}, which predicts

%\begin{equation}
%\label{dgwilson}
%    S_k^{\rm DF}=-2N\,\Big[\frac{k\sqrt{\lambda}}{4N}\,\sqrt{1+\frac{k^2\lambda}{16N^2}}+{\rm arcsinh}\Big(\frac{k\sqrt{\lambda}}{4N}\Big)\Big]\,.
%\end{equation}
%
\begin{equation}
\label{dgwilson}
    S_k^{\rm DF}=-2N\,\Big[\tilde\kappa \,\sqrt{1+\tilde\kappa^2}+{\rm arcsinh}\big(\tilde\kappa\big)\Big]\, ,
\end{equation}
where
\begin{equation}
    \tilde\kappa = \frac{k\sqrt{\lambda}}{4N} \qquad \leadsto \qquad \tilde\kappa ^2=\kappa \ \frac{k}{16N}\ .
\end{equation}
This formula is obtained by using the fact that the Wilson loop in the $k$-symmetric representation corresponds to D3 branes with $k$ units of electric flux, or, equivalently, with
$k$ non-interacting strings. Note that it assumes fixed $\tilde{\kappa}$ and $N\gg k\gg 1$. In this regime, in a $1/N$ expansion, both the $k$-symmetric and $k$-wound loops give the same result \eqref{dgwilson} \cite{Gomis:2006sb,Hartnoll:2006is}.\footnote{In fact, even the first $1/N$ correction agrees between both for $N\gg k\gg 1$ at fixed small $\frac{\sqrt{\lambda}\,k}{N}$ \cite{Buchbinder:2014nia}. 
}

A priori, it is not guaranteed that  gauge theory and supergravity results should match, since the formula (\ref{result3}) requires
$k\gg N\gg 1$, a regime where back-reaction effects on the $AdS_5\times S^5$ geometry could be important.
Nonetheless, we can examine the holographic formula (\ref{dgwilson}) in the limit $\tilde\kappa\gg 1$, although aware of this possible limitation.
Expanding (\ref{dgwilson})   for  $\tilde\kappa\gg 1$
and writing the result in terms of $\kappa$, one finds

\begin{equation}
    S_k^{DF}\sim -\frac{k\kappa}{8}-N \log\Big(\frac{k\,\kappa}{4N}\Big)-N\,.
\end{equation}
Notably, this coincides with the gauge theory result \eqref{result3}.

\subsection{The $SU(N)$ case}

In order to compute the VEV of the Wilson loop $W_k$ in the theory with gauge group $SU(N)$ it is convenient to begin with the $U(N)$ case, where

\begin{equation}
    \langle W_k\rangle=\frac{1}{Z_{U(N)}} \int d^Na\,\prod_{k<l}(a_k-a_l)^2\,e^{-\frac{8\pi^2}{g^2}\sum_{m=1}^N a_m^2}\frac{e^{2\pi (k+N-1)\,a_N}}{\prod_{j\ne N} \big(e^{2\pi a_N}-e^{2\pi a_j}\big)}\,. 
\end{equation}
Using that

\begin{equation}
    \sum_{i=1}^Na_i^2=\sum_{i=1}^{N} \hat{a}_i^2+\frac{1}{N}(\sum_{i=1}^Na_i)^2\,,\qquad \hat{a}_i \equiv a_i-\frac{1}{N}\,\sum_{i=1}^N a_i\ ,
\end{equation}
$ \langle W_k\rangle $ can be written as

\begin{equation}
    \langle W_k\rangle=\frac{1}{Z_{U(N)}} \int d^Na\,\prod_{k<l}(\hat{a}_k-\hat{a}_l)^2\,e^{-\frac{8\pi^2}{g^2}\sum_{m=1}^N \hat{a}_m^2}\,e^{-\frac{8\pi^2 N}{g^2}\,x^2-2\pi k x}\frac{e^{2\pi (k+N-1)\,\hat{a}_N}}{\prod_{j\ne N} \big(e^{2\pi \hat{a}_N}-e^{2\pi \hat{a}_j}\big)}\,,
\end{equation}
where $x=\frac{1}{N}\sum_{i=1}^Na_i$ and
 $\sum_{i=1}^N\hat{a}_i=0$. 
 %We can relax this constraint at the expense of introducing a $\delta(\sum_{i=1}^N\hat{a}_i)$. 
 This leads to

\begin{eqnarray}
    \langle W_k\rangle &=&\frac{1}{Z_{U(N)}} \int d^N\hat{a}\,\prod_{k<l}(\hat{a}_k-\hat{a}_l)^2\,e^{-\frac{8\pi^2}{g^2}\sum_{m=1}^N \hat{a}_m^2}\,\frac{e^{2\pi (k+N-1)\,\hat{a}_N}}{\prod_{j\ne N} \big(e^{2\pi \hat{a}_N}-e^{2\pi \hat{a}_j}\big)}\,\delta(\sum_{i=1}^N\hat{a}_i)\,
    \nonumber\\
    &\times& \Big(\int dx \,e^{-\frac{8\pi^2 N}{g^2}\,x^2-2\pi k x}\Big)\, .
\end{eqnarray}
Thus

\begin{equation}
\label{WSUWU}
    \langle W_k\rangle_{U(N)}=\left(\frac{Z_{SU(N)}}{Z_{U(N)}}\,\int dx\, e^{-\frac{8\pi^2 N}{g^2}\,x^2-2\pi k x}\right)\,\langle W_k\rangle_{SU(N)}\,.
\end{equation}
Computing the integral, we find

\begin{equation}
\label{UvsSU}
    \langle W_k\rangle_{U(N)}=e^{\frac{\kappa\,k}{8\,N}}\,\langle W_k\rangle_{SU(N)}\,.
\end{equation}
One can check this formula in two different ways: setting $N=2$ we recover our  result above with $k$ arbitrary; setting $k=1$ and arbitrary $N$ we recover a formula given in \cite{Drukker:2000rr}.

It should be noted that even for arbitrarily large $N$ (while still much smaller than $k$ so that the  approximation holds), the loops in the $SU(N)$ and $U(N)$ theory do not coincide. To understand this, let us look to the prefactor in \eqref{WSUWU}. It can be written as follows

\begin{equation}
   \frac{Z_{SU(N)}}{Z_{U(N)}}\,\int dx\, e^{-\frac{8\pi^2 N}{g^2}\,x^2-2\pi k x}=\frac{\int da\, e^{-\frac{8\pi^2}{g^2}\,a^2-2\pi \frac{k}{\sqrt{N}} a}}{\int da\, e^{-\frac{8\pi^2}{g^2}\,a^2}}\ .
\end{equation}
Thus, the prefactor corresponds to the contribution of a  Wilson loop of the $U(1)$ theory with charge $\frac{k}{\sqrt{N}}$.\footnote{Note that the generator in $U(1)\in U(N)$ must be normalized with a $N^{-\frac{1}{2}}$ so that ${\rm Tr}T^2\sim 1$.} As this is much larger than 1 in our limit, we see that the ${\rm Sym}^k(\Box)$ Wilson loop has an overall factor originating from the extra $U(1)$ that cannot be neglected even if $N\gg 1$. As a consequence, the loops in the $U/SU$ theories are different already to leading order in the large $k$ expansion. 
This produces an intriguing mismatch with the holographic formula when the gauge group is $SU(N)$. One may therefore view $\langle W_k\rangle$ with $k\gg N$ as an  observable that distinguishes the large $N$ $SU(N)$ theory from the large $N$ $U(N)$ theory.

%%%%%%%%%%%%%%%%%%%%%%%%%%%%%%%%%%%%%%%%%%%%
\section{Correlation functions of  ${\rm Sym}^k(\Box)$ Wilson loops with CPO's}\label{CPOsW}

As discussed in \cite{Gerchkovitz:2016gxx,Rodriguez-Gomez:2016ijh,Rodriguez-Gomez:2016cem}, computing correlation functions in $\mathbb{R}^4$ from the matrix model involves a conformal map from $\mathbb{S}
^4$ into $\mathbb{R}^4$.  The $R$-charge is not conserved in correlation functions in $\mathbb{S}^4$. This is possible because
the theory on $\mathbb{S}^4$ breaks the $U(1)_R$ symmetry.
These mixtures are a reflection of the conformal anomaly in $\mathbb{S}^4$. The four-sphere introduces a  scale, the radius,
which leads to a mixture of operators of different dimensions.
The standard correlation functions of the theory in flat space can be recovered by a Gram-Schmidt procedure introduced in \cite{Gerchkovitz:2016gxx} (see appendix \ref{CPOs} for a lightning review), by which one can find orthogonalized operators in the sphere matrix model which map to the $\mathbb{R}^4$ operators. As shown in \cite{Rodriguez-Gomez:2016cem}, the obtained operators can then be used to compute correlation functions of the CPO's with circular Wilson loops (for a closely related approach, see  \cite{Billo:2018oog}). We are now interested in applying this method to the correlator of CPO's with Wilson loops in the $k$-symmetric representation.
As a first step, consider, in the $\mathbb{S}^4$ matrix model, the correlation function 

\begin{eqnarray}
    \langle {\rm Tr}\phi^{n_1}\cdots {\rm Tr}\phi^{n_m}\,W_k\rangle&=&\frac{1}{Z_{U(N)}} \int d^Na\,\prod_{k<l}(a_k-a_l)^2\,Z_{\rm 1-loop}\,Z_{\rm inst}\,e^{-\frac{8\pi^2}{g^2}\sum_{m=1}^N a_m^2}\nonumber \\ &\times & \frac{e^{2\pi (N-1)\,a_N+2k\pi a_N}}{\prod_{j\ne N} \big(e^{2\pi a_N}-e^{2\pi a_j}\big)}\,\Big(\sum_{i=1}^N a_i^{n_1}\Big)\cdots \Big(\sum_{i=1}^Na_i^{n_m}\Big)\,. 
\end{eqnarray}
For large $k$, exactly the same argument as in section 4 can be used: the $a_i$'s with $i\neq N$ will only contribute in the integration region very close to zero whereas the main contribution of the integral over $a_N$ will come from the region $a_N\sim\kappa/8\pi$. In the double-scaling limit,
this approximation becomes exact.
Therefore, we are led to

\begin{equation}
     \langle {\rm Tr}\phi^{n_1}\cdots {\rm Tr}\phi^{n_m}\,W_k\rangle=\frac{Z_{U(N-1)}}{Z_{U(N)}}\,e^{\frac{k\kappa}{8}\,(1+\frac{N-1}{k})}\, \int da_N\,\Big(\frac{a_N^2}{e^{2\pi a_N}-1}\Big)^{N-1}\,a_N^{n_1+\cdots+n_m}\,e^{-k\frac{8\pi^2}{\kappa}(a_N-\frac{\kappa}{8\pi})^2}\,. 
\end{equation}
This gives the simple result

\begin{equation}
     \langle {\rm Tr}\phi^{n_1}\cdots {\rm Tr}\phi^{n_m}\,W_k\rangle=\Big(\frac{\kappa}{8\pi}\Big)^{n_1+\cdots+n_m}\,\langle W_k\rangle\,. 
\end{equation}
We now consider the orthogonalization process.
The first few  operators, using Gram-Schmidt, are

\begin{equation}
    {\cal O }_1={\rm Tr}\phi-\frac{\langle {\rm Tr}\phi\rangle}{\langle \unity\rangle}\,\unity\,,\qquad
    {\cal O }_2={\rm Tr}\phi^2-\frac{\langle {\rm Tr}\phi^2\rangle}{\langle \unity\rangle}\,\unity\,\qquad \cdots .
\end{equation}
Here ${\cal O}_\Delta$ refers to the $\mathbb{R}^4$ operator. Then

\begin{equation}
    \langle {\cal O}_1\,W_k\rangle=\Big[\frac{\kappa}{8\pi}-\frac{\langle {\rm Tr}\phi\rangle}{\langle \unity\rangle}\Big]\,\langle W_k\rangle\,,\qquad
    \langle {\cal O}_2\,W_k\rangle=\Big[\Big(\frac{\kappa}{8\pi}\Big)^2-\frac{\langle {\rm Tr}\phi^2\rangle}{\langle \unity\rangle}\Big]\,\langle W_k\rangle\,\qquad \cdots.
\end{equation}
Now, the VEV's $\frac{\langle {\rm Tr}\phi^n\rangle}{\langle\unity\rangle}$ in the matrix model without the insertion of $W_k$ is proportional to $g^n$. Expressed in terms of $k$, this is equal to
$\kappa^{\frac{n}{2}}\,k^{-n}$.
Thus, in limit of large $k$ with fixed $\kappa$,  this is suppressed.

It is clear that this argument will hold true for arbitrary operators: the $\mathbb{S}^4$ mixing with lower operators arises through terms with $\langle {\rm Tr}\phi^{\Delta_1}\cdots {\rm Tr}\phi^{\Delta_l}\rangle$, which have no insertion of $W_k$ and are similarly suppressed in the large $k$ limit (here we assume $\Delta_i\ll k$).  Hence we obtain the remarkably simple formula

\begin{equation}
\label{WO}
    \langle {\cal O}_{\Delta}\,W_k\rangle=\Big(\frac{\kappa}{8\pi}\Big)^{\Delta}\,\langle W_k\rangle\,,
\end{equation}
for any CPO of dimension $\Delta$. As a sanity check of this formula, one may compute

\begin{equation}
\label{WOdos}
    \langle {\cal O}_2 \,W_k\rangle=\frac{\partial}{\partial x}(Z_N\langle W_k\rangle)=\Big(\frac{\kappa}{8\pi}\Big)^{2}\,\langle W_k\rangle\,,\qquad x=-\frac{8\pi^2}{g^2}\,;
\end{equation}
up to $\frac{1}{k}$ corrections.

It should be noted that ${\cal O}_{\Delta}$ in \eqref{WO} is normalized such that (see \textit{e.g.} \cite{Rodriguez-Gomez:2016ijh})

\begin{equation}
    \langle \mathcal{O}_{\Delta}(x)\,\mathcal{O}_{\Delta}(0)\rangle=\frac{C_{\Delta}}{|x|^{2\Delta}}\, ,\qquad C_{\Delta}\equiv \frac{\Delta\,\lambda^{\Delta}}{(2\pi)^{2\Delta}}\,.
\end{equation}
Introduce now $\hat{\mathcal{O}}_{\Delta}=C_{\Delta}^{-\frac{1}{2}}\,\mathcal{O}_{\Delta}$, so that the CPO's have a ``canonical" normalization
$$
\langle \hat{\mathcal{O}}_{\Delta}(x)\,\hat{\mathcal{O}}_{\Delta}(0)\rangle=\frac{1}{|x|^{2\Delta}}\ .
$$
Then, we obtain the formula

\begin{equation}
\label{predw}
    \langle {\cal \hat{O}}_{\Delta}\,W_k\rangle=\frac{1}{\sqrt{\Delta}}\,\Big(\frac{k\kappa}{16 N}\Big)^{\frac{\Delta}{2}}\,\langle W_k\rangle\,,
\end{equation}
This can be compared with the similar formula
derived for the multiply wound fundamental Wilson loop in \cite{Okuyama:2006jc,Giombi:2006de}. For large $\lambda$, 
the $k$-wound fundamental Wilson loop should give rise to the same correlation functions as the Wilson loop in the $k$-symmetric representation. Indeed,
taking $\lambda\gg 1$ in the formulas of \cite{Giombi:2006de}, we find agreement with (\ref{predw}).\footnote{See  (5.9) and (5.10) in \cite{Giombi:2006de}, taking into account the different definition of the parameter $\kappa $. In \cite{Giombi:2006de}, $\kappa \equiv \frac{k \sqrt{\lambda}}{4N}$, which, in our notation, corresponds to $ \frac{\sqrt{k}\, \sqrt{\kappa}}{4\sqrt{N}}$\, .}

\section{$\mathcal{N}=2^*$ theory}\label{N=2*}

The method for computing $\langle W_k\rangle$ at large $k$, fixed $N$, used in the previous sections for the $\mathcal{N}=4$ theory can be
extended to any $\mathcal{N}=2$ theory.
As an example, here we shall consider the $\mathcal{N}=2^*$ theory, defined as usual by giving a mass $M$ to the
hypermultiplet.
In this theory the coupling constant does not run and it is a parameter that characterizes the theory.
The $\mathcal{N}=2^*$ theory was thoroughly studied using supersymmetric localization in a series of papers, starting with \cite{Russo:2013qaa,Russo:2013kea}. It was found that, in the decompactification limit where the radius $R$ of $\mathbb{S}^4$ goes to infinity,
the large $N$ theory undergoes an infinite number of phase transitions as the 't Hooft coupling is varied from 0 to infinity.

The first phase transition occurs at $\lambda_c\cong 35.4$; then there is a second phase transition occurring at $\lambda \cong 83$, a third phase transition at $\lambda \cong 150$, followed by 
an infinite sequence of phase transitions, which for large $\lambda$ occur at critical values $\lambda\approx n^2\pi^2$,
with integer $n\gg 1$.
The $SU(2)$ theory does not have phase transitions \cite{Russo:2014nka}. However, for  any finite $N>2$, there is evidence that at least the first transition   must occur at the same
$\lambda_c\cong 35.4$ \cite{Hollowood:2015oma}. Similar phase transitions
are generically expected in massive $\mathcal{N}=2$ theories. Another example is provided by massive SQED with $N_f<2N$ at large $N$ \cite{Russo:2013kea} as well as for any $SU(N)$ gauge group with $N\geq 2$ \cite{Russo:2014nka,Russo:2015vva,Russo:2019ipg}.

In the limit considered in this paper, with $g\to 0$ and $N$ fixed, the 't Hooft coupling $\lambda=g^2 N$ vanishes. Therefore the above phase transitions do not occur. An interesting question is whether there could still be phase transitions for the one-dimensional defect theory defined by the insertion of $W_k$, at critical values of the parameter
$\kappa\equiv k g^2$.

The VEV of the Wilson loop for gauge group $U(N)$ is now
\begin{equation}
\label{wkNstar}
    \langle W_k\rangle _{\mathcal{N}=2^*}=\frac{1}{Z_{\mathcal{N}=2^*}} \int d^Na\,\prod_{k<l}(a_k-a_l)^2\,Z_{\rm 1-loop}\,e^{-\frac{8\pi^2}{g^2}\sum_{m=1}^N a_m^2}\frac{e^{2\pi (k+N-1)\,a_N}}{\prod_{j\ne N}\big( e^{2\pi a_N}-e^{2\pi a_j}\big)}\,.
\end{equation}
where we have omitted the instanton factor as this is suppressed in the double-scaling limit \eqref{limitedoble}. The one-loop factor is given by \eqref{1loopN2*}. 
The integrals can be computed by following the same procedure as in the $\mathcal{N}=4$ case.
We first separate the factors with $a_N$ dependence.
Then \eqref{wkNstar} becomes

\begin{eqnarray}
    \langle W_k\rangle _{\mathcal{N}=2^*}&=&\frac{1}{Z_{\mathcal{N}=2^*}} \int d^Na\, \prod_{i<j}^{N-1}\frac{(a_i-a_j)^2H(a_i-a_j)^2}{H(a_i-a_j+M)H(a_i-a_j-M)}\,e^{-\frac{8\pi^2}{g^2}\sum_{m=1}^{N-1} a_m^2}\nonumber \\ && \hspace{-2cm} e^{\frac{k^2g^2}{8}\big(1+\frac{N-1}{k}\big)^2}\,\prod_{i=1}^{N-1}\frac{(a_i-a_N)^2 H(a_i-a_N)^2}{H(a_i-a_N+M)H(a_i-a_N-M)}\,\frac{e^{-\frac{8\pi^2 }{g^2}(a_N-a_N^{* })^2}}{\prod_{j\ne N} \big(e^{2\pi a_N}-e^{2\pi a_j}\big)}\,,
\end{eqnarray}
where $a_N^{* }=\kappa/(8\pi)$ just  as in the $\mathcal{N}=4$ case.
Next, we introduce new integration variables $x_i=a_i/g$, $i=1,...,N-1$ and expand 
the factors in the integrand depending on $x_i$ in powers of $g$.
In the limit $g\to 0$, we are left with the leading term, which, in terms of the original variables $a_i$, reads

\begin{eqnarray}
    \langle W_k\rangle _{\mathcal{N}=2^*}&=&\frac{e^{\frac{k\kappa}{8}\big(1+\frac{N-1}{k}\big)^2}}{Z_{\mathcal{N}=2^*}} \int d^{N-1}a\, \prod_{i<j}^{N-1}\frac{(a_i-a_j)^2H(a_i-a_j)^2}{H(a_i-a_j+M)H(a_i-a_j-M)}\,e^{-\frac{8\pi^2k}{\kappa}\sum_{m=1}^{N-1} a_m^2}\nonumber \\ && \hspace{-2cm} \,\int da_N\,\Big(\frac{H(a_N)^2}{H(a_N+M)H(a_N-M)}\Big)^{N-1}\,\Big(\frac{a_N^2}{e^{2\pi a_N}-1}\Big)^{N-1}\,e^{-k\frac{8\pi^2 }{\kappa}(a_N-a_N^{* })^2}\, ,
\end{eqnarray}
where $\kappa\equiv g^2k$.
In the first line we recognize the partition function for the $\mathcal{N}=2^*$ theory with gauge group $U(N-1)$. In turn, the integral in the second line can be easily done through saddle point. Collecting all factors

\begin{equation}
    \langle W_k\rangle _{\mathcal{N}=2^*}=\frac{Z^{U(N-1)}_{\mathcal{N}=2^*}}{Z^{U(N-1)}_{\mathcal{N}=4}}\,\frac{Z^{U(N)}_{\mathcal{N}=4}}{Z^{U(N)}_{\mathcal{N}=2^*}}  \Big(\frac{H(a^{* }_N)^2}{H(a^{* }_N+M)H(a^{* }_N-M)}\Big)^{N-1}  \langle W_k\rangle_{\mathcal{N}=4} \,.
\end{equation}
Similarly, in the $g\to 0$ limit, one has

\begin{equation}
   Z^{U(N)}_{\mathcal{N}=2^*}= \prod_{i<j}^{N}\frac{1}{H(M)^2}\, \int d^Na\,\prod_{i<j}^{N}(a_i-a_j)^2\,e^{-\frac{8\pi^2k}{\kappa}\sum_{m=1}^{N} a_m^2}= \frac{Z^{U(N)}_{\mathcal{N}=4} }{H(M)^{N(N-1)} }\,.
\end{equation}
Thus we finally obtain

\begin{equation}
\label{wkresulta}
    \langle W_k\rangle _{\mathcal{N}=2^*}= \Big(\frac{H(a^{* }_N)^2\,H(M)^2}{H(a^{* }_N+M)H(a^{* }_N-M)}\Big)^{N-1}  \langle W_k\rangle _{\mathcal{N}=4} \,.
\end{equation}
This is the main result of this section.

\medskip

Note that $f\equiv -\log \langle W_k \rangle_{\mathcal{N}=2^*}  $ represents the  free energy of the one-dimensional defect theory  on $S^1$. An interesting question regards the behavior of $f$  as a function of $\kappa$.
In particular, whether $f(\kappa)$ exhibits non-analytic behavior in the infinite volume theory.

The decompactification limit corresponds to sending $MR\to \infty$, $R$ being the radius of the four-sphere.
The dependence on $R$ is restored by $M\to MR$, $a_N^*\to Ra_N^*$, {\it i.e.} $a_N^*=\kappa/(8\pi R)$.
The expansion of the function $H(x)$ for large argument is derived from the asymptotic expansion of the Barnes G-function. One finds
\begin{equation}
    \log H(x) = -\frac12 x^2\log x^2 +\left(\frac12 -\gamma\right) x^2+ {\cal O}(\log x^2 )\ .
\end{equation}
Thus
\begin{eqnarray}
    \log  \langle W_k \rangle_{\mathcal{N}=2^*}&\approx &
     \log  \langle W_k \rangle_{\mathcal{N}=4} +2(N-1)\log H(a^{* }_N)
     - \frac12 (N-1)R^2 \Big[2M^2\log (MR)^2 
    \nonumber\\
    &-&(M-a_N^*)^2\log (M-a_N^*)^2R^2 -(M+a_N^*)^2\log (M+a_N^*)^2R^2 \Big]
     \nonumber\\
     &+&(1-2\gamma) (N-1)(Ra_N^*)^2
    \,.
\end{eqnarray}

A potential non-analytic behavior is at $a_N^*=\pm M$, that is, $\kappa=8\pi M R$. 
Since $R\to\infty $, this point is not reached for any finite $\kappa $.
Indeed, at large $R$, $a_*$ is small compared to $M$, so in $\mathbb{R}^4$ we effectively have
\begin{equation}
    \log  \langle W_k \rangle_{\mathcal{N}=2^*}\ \longrightarrow\ 
     \log  \langle W_k \rangle_{\mathcal{N}=4} + (N-1)\left(2 \log H(\frac{\kappa}{8\pi})
    + \frac{\kappa^2}{32\pi^2}\Big[2-\gamma+\log\big(  MR\big) \Big]\right) \ .
\end{equation}
where $ \log  \langle W_k \rangle_{\mathcal{N}=4} $ is the  function
of $\kappa $ given in \eqref{logresult}. Note the logarithmic infrared divergence, which is due to the presence of massless particles.
The resulting ``free energy" $f$ of the defect theory is a smooth function of $\kappa$.

%*Observamos una divergencia infraroja, $\propto \kappa^2\log(MR)$. Que hacemos con eso?

%Consider
%\begin{equation}
%    \partial_\kappa f= \frac{k}{8}+(N-1)\left( \frac{1}{\kappa} -\frac{e^{-\kappa/4}}{4(1-e^{-\kappa/4}) } +\frac{\kappa}{16\pi^2}(1-\log\Big( \frac{\kappa}{8\pi MR})\right)
%\end{equation}
%Se podria tratar de entender mejor. Si lo ploteamos, la primera pregunta es si
%$\log<W>$ se anula para algun $\kappa $ (reemplazando $MR$ por una escala IR finita, aunque no se que significa).  Nota que $\log<W>=0$ es como tension de cuerda nula.

%Esta claro que $f$ es smooth, pero podria tener informacion interesante.
\smallskip

In conclusion, we have computed   $\langle W_k \rangle$ in the large $k$ limit for the $\mathcal{N}=2^*$ $U(N)$ theory on $\mathbb{S}^4$.
The resulting expression  \eqref{wkresulta} exhibits an interesting interplay between the two scales $a_N^*=\kappa/(8\pi R)$ and $M$.
At infinite volume, one has $a_*/M\to 0$ and the VEV of the loop
becomes identical to the case of the $\mathcal{N}=4$ theory computed in previous sections.
Consequently, the associated free energy $f(\kappa )=-\log \langle W_k \rangle_{\mathcal{N}=2^*}$ of the defect theory  does not exhibit non-analytic behavior.

\section{Discussion}\label{Discussion}

In this note we have studied circular Wilson loops in the  $k$-symmetric representation in 4d $\mathcal{N}=2$ theories with gauge group $U(N)$ or $SU(N)$ using a double-scaling limit. This limit gives rise to  exact results for any finite $N$,
which include all perturbative contributions.
Gauge instanton contributions exponentially vanish in the limit.  The VEV of the Wilson loop  contains contributions from the 1-loop determinant, which is generically expressed in terms of Barnes G-functions (see \eqref{wkresulta} for the case of the $\mathcal{N}=2^{* }$ theory). The resulting formula represents the resummation of infinitely many Feynman diagrams in standard perturbation theory.

The  limit studied here corresponds to $k\rightarrow \infty$ while $\kappa\equiv g^2k$ fixed at finite $N$. Effectively, this implies $k\gg N$. Clearly, this is different from taking the large $N$ limit at fixed $k/N$, but there is a region of overlapping.
%, since the latter implicitly assumes $k\ll N$.
Indeed, one can  study the large $N$ behavior of the  expressions obtained by the double-scaling limit \eqref{limitedoble}, as long as $\frac{N}{k}\ll 1$. 
For the $U(N)$ $\mathcal{N}=4$ SYM theory, we have found agreement with the most familiar large $N$ limit at fixed $k/N$, which has been studied in the literature, both from the QFT matrix model perspective and holographically. 

 An
interesting aspect of the limit at fixed $N$ discussed here is that it distinguishes between the $U(N)$ and the $SU(N)$ theory, even if $N\gg 1 $; see \eqref{UvsSU}. 
This result opens the door to new precision tests of holography, as it can be used to probe holographic properties of the diagonal $U(1)$ in $U(N)$. In particular, it would be very interesting to see if
the $SU(N)$ result, including the (leading!) prefactor in \eqref{UvsSU}, can be recovered using holography, upon adding the suitable boundary conditions. The idea is as follows.
%(along the lines of \cite{Hofman:2017vwr}).  
Recall that the global properties of the gauge group are encoded in the topological sector of Type IIB supergravity on $AdS_5$ after reduction on the $\mathbb{S}^5$. This results on a BF theory in the bulk with 
$$
S=(2\pi)^{-1}\,N\int_{AdS_5}C_2\wedge dB_2\ ,
$$ 
where $C_2$ and $B_2$ are the RR and NS 2-form potentials respectively. As discussed in \cite{Hofman:2017vwr}, this action has to be supplemented with appropriate boundary terms to impose the desired boundary conditions that define the $U(N)$ or $SU(N)$ theory (or other quotients by the center). To make contact with our discussion, recall that the Wilson loop in the $k$-symmetric representation is
holographically represented by a D3 brane with electric flux dissolving $k$ fundamental strings \cite{Drukker:2005kx}. This is a source for the RR 2-form potential $C_2$ entering in the BF topological theory controlling the global properties of the gauge group. Therefore, we expect that one can match the $SU(N)$ gauge theory result by adding suitable boundary terms. While these boundary contributions should be negligible for $\frac{k}{N}\ll 1$, they should be relevant in the limit $\frac{k}{N}\gg 1$.

The formula \eqref{logresult} for $\log\langle W_k\rangle $ exhibits some features that are inherent to the $k$-symmetric representation, such as the 
presence of an infinite series of exponentially  terms
of the form $e^{-n \kappa/4}$ in a large $\kappa$ expansion.
These contributions are associated with the massive particles at the point in moduli space that dominates the path integral, {\it i.e.} at the saddle point \cite{Hellerman:2021duh}. One can understand this feature from the spectrum.
In general, for any ${\cal N}=2 $ SYM, the spectrum contains
massive vector multiplets with masses
\begin{equation}
    M^V_{ij}=|a_i-a_j|\ .
\end{equation}
At the saddle point, there are $N-1$ vector multiplets
with masses
\begin{equation}
    M^V_{ij}=|a_*|\ .
\end{equation}
In addition, there are massive hypermultiplets at the saddle point. The masses depend on the case. For the ${\cal N}=4$ theory, because of supersymmetry, their masses
coincide with the above mass spectrum of the vector multiplets.
The action of a particle with mass $m= |a_*|=\kappa/8\pi$ circulating
around the equator of $\mathbb{S}^4$ is $S=2\pi m =\kappa/4$.
In addition, there are BPS electric particles of masses $n |a_*|$ corresponding to BPS bound states. Thus one expects  an infinite series of contributions $e^{-n \kappa/4}$, which indeed appear in the formula for $\log\langle W_k\rangle $, multiplied by $N-1$, which is the correct degeneracy.

Alternatively, one can interpret $\langle W_k\rangle$ in terms of degrees of freedom on the 1d defect theory  on $\mathbb{S}^1$. 
From \eqref{logresult}, we have

\begin{equation}
\label{wboson}
    \langle W_k\rangle =\frac{1}{N!}\, \, e^{\pi m k}\, \, \, \left(Z_m\right)^{N-1} \ ,\qquad Z_m =\frac{2\pi mk}{1-e^{-2\pi m}} \,,\qquad m\equiv \frac{\kappa}{8\pi }\ ,
\end{equation}
where $(Z_m)^{N-1}$ is to be interpreted as the partition function for
$N-1$ bosons on the one-dimensional defect $\mathbb{S}^1$
(see similar discussions in \cite{Gomis:2006im,Hoyos:2018jky,Beccaria:2022bcr}).

Here we have focused on $\mathcal{N}=4$ SYM with unitary gauge group and its massive deformation --the $\mathcal{N}=2^{* }$ theory--, even though a similar limit exists for other  $\mathcal{N}=2$ theories and other gauge groups (see \cite{Cuomo:2022xgw} for a study of $SU(2)$ SQCD). In particular, it would be interesting to study the double-scaling limit for Wilson loops in
symmetric representations in quiver gauge theories, also including  correlation functions involving CPO's. Such correlation functions can be computed from a matrix model \cite{Billo:2021rdb} and it would be interesting to study its potential applications to Wilson loops. 

A more ambitious goal would be to study Wilson loops in large representations in non-supersymmetric Yang-Mills theory, by a similar  double-scaling limit at fixed $N$  in the
UV, where $g$ is small. Obtaining exact results for these observables could reveal  interesting new
features  of Yang-Mills theory.

\section*{Acknowledgements}

We would like to thank Arkady Tseytlin for useful discussions. D.R-G is partially supported by the Spanish government grant MINECO-16-FPA2015-63667-P. He also acknowledges support from the Principado de Asturias through the grant FC-GRUPIN-IDI/2018/000174 J.G.R. acknowledges financial support from projects  MINECO
grant PID2019-105614GB-C21, and  from the State Agency for Research of the Spanish Ministry of Science and Innovation through the “Unit of Excellence María de Maeztu 2020-2023” (CEX2019-000918-M).

\begin{appendix}

\section{Correlation functions for CPO's using localization}\label{CPOs}

In this appendix we briefly review the computation of correlation functions for chiral primary operators (CPO's) through supersymmetric localization following the construction in \cite{Gerchkovitz:2016gxx}. Recall that, in Lagrangian theories, CPO's correspond to operators $\mathcal{O}_{\Delta}$ made out of scalar fields in vector multiplets. Conformal invariance dictates that  2-point functions in $\mathbb{R}^4$ must be of the form 

\begin{equation}
    \langle \mathcal{O}_{\Delta_1}(x)\,\mathcal{O}_{\Delta_2}(0)\rangle=\frac{g_{\Delta_1,\Delta_2}}{|x|^2}\,\delta_{\Delta_1,\Delta_2}\,.
\end{equation}
Thus, all the non-trivial information resides in the Zamolodchikov metric $g_{\Delta_1,\Delta_2}$, which in general depends on the marginal couplings. To further proceed, one notes that one can extract $g_{\Delta_1,\Delta_2}$ by considering

\begin{equation}
    4^{\Delta_1}\,\langle\,  \lim_{|x|\rightarrow \infty} (1+\frac{|x|^2}{4})^{\Delta_1} \mathcal{O}_{\Delta_1}(x)\,\mathcal{O}_{\Delta_2}(0)\, \rangle=g_{\Delta_1,\Delta_2}\,\delta_{\Delta_1,\Delta_2}\,.
\end{equation}
We recognize the conformal factor mapping the plane into the sphere.
Therefore, the relevant information of correlation function in  $\mathbb{R}^4$
can be computed through a correlation function in the sphere, where the CPO's are inserted in the North/South poles:

\begin{equation}
\label{arfe}
    \langle  \mathcal{O}_{\Delta_1}(N)\,\mathcal{O}_{\Delta_2}(S)\rangle_{\mathbb{S}^4}=4^{-\Delta_1}\,g_{\Delta_1,\Delta_2}\,\delta_{\Delta_1,\Delta_2}\,;
\end{equation}
As shown in \cite{Gerchkovitz:2016gxx}, due to the supersymmetric properties of the CPO's, this correlation function coincides with the one corresponding to the insertion of the integrated (super) field. 
This justifies the use of supersymmetric localization to compute the sphere 2-point functions through the corresponding matrix model. 
The construction is as follows. 
We begin by defining operators ${O}_{\Delta}$ on $\mathbb{S}^4$ as
$$
{O}_{\Delta}\equiv {\rm Tr}\phi^{p_1}\cdots {\rm Tr}\phi^{p_P}\ ,\qquad \sum_{i=1}^P p_i=\Delta\ ,
$$
where $\phi $ is the adjoint scalar field of the $\mathcal{N}=1$ vector multiplet.
Localization reduces the functional integral of $\mathcal{N}=2$ $U(N)$ gauge theories  to a finite $N$-dimensional integral
over the moduli space parametrized by the diagonal expectation
value $\phi ={\rm diag}(a_1,...,a_N) $ \cite{Pestun:2007rz}.
Correlation functions of  operators ${O}_{\Delta_i}$ can then be computed by 
\begin{eqnarray}
    \langle  {O}_{\Delta_1}(N)\,{O}_{\Delta_2}(S)\rangle_{S^4}&=&4^{-\Delta_1}\,\int d^Na \prod_{i<j}^N (a_i-a_j)^2\,e^{-\frac{8\pi^2}{g^2}\sum_{i=1}^Na_i^2} \\ \nonumber && \Big[\Big(\sum_{i=1}^N a_i^{p_1}\Big)\cdots \Big(\sum_{i=1}^N a_i^{p_P}\Big)\Big]\,\Big[\Big(\sum_{i=1}^N a_i^{q_1}\Big)\cdots \Big(\sum_{i=1}^N a_i^{q_Q}\Big)\Big]\,.
\end{eqnarray}
This correlation function is non-zero for operators ${O}_{\Delta_i}$ of different dimensions, which shows that the ${O}_{\Delta_i}$ cannot be identified with the operators $\mathcal{O}_{\Delta_i}$ in \eqref{arfe}. 
The underlying reason for
why the correlation functions of the operators of different dimensions can be nonzero is the conformal anomaly of the theory on the four-sphere.
The sphere has an intrinsic scale --its radius $R$. When mapping $\mathbb{R}^4$ operators to $\mathbb{S}^4$ operators, operators of different dimensions get mixed. On general grounds, one obtains a relation of the form

\begin{equation}
\label{mixings}
    \mathcal{O}_{\Delta}^{\mathbb{R}^4}=O_{\Delta}^{\mathbb{S}^4}+\frac{\alpha^{\Delta}_2}{R^2}O_{\Delta-2}^{\mathbb{S}^4}+\frac{\alpha^{\Delta}_4}{R^2}O_{\Delta-4}^{\mathbb{S}^4}+\cdots\,.
\end{equation}
The key insight of \cite{Gerchkovitz:2016gxx} is that 
the standard two-point correlation functions in $\mathbb{R}^4$ proportional to $\delta_{\Delta_1,\Delta_2}$
can be recovered by a Gram-Schmidt orthogonalization procedure. For instance, for $U(2)$ $\mathcal{N}=4$, on $\mathbb{R}^4$ one has the CPO's $\mathcal{O}^{\mathbb{R}^4}_1={\rm Tr}\phi$ and $\mathcal{O}^{\mathbb{R}^4}_2={\rm Tr}\phi^2$ (at larger dimensions there are only multitraces), while in the sphere the relevant operators are $\unity$, ${\rm Tr}\phi_{\mathbb{S}^4}$ and ${\rm Tr}\phi^2_{\mathbb{S}^4}$. One then has

\begin{eqnarray}
&&\mathcal{O}_1^{\mathbb{R}^4}\rightarrow {\rm Tr}\phi_{\mathbb{S}^4}-\frac{\langle {\rm Tr}\phi_{\mathbb{S}^4}\rangle}{\langle \unity\rangle}\,\unity\,,\\
&& \mathcal{O}_2^{\mathbb{R}^4}\rightarrow {\rm Tr}\phi_{\mathbb{S}^4}^2- \frac{\langle {\rm Tr}\phi^2_{\mathbb{S}^4}\rangle}{\langle \unity\rangle}\,\unity\,;
\end{eqnarray}
where the $\langle \cdot\rangle$ is to be computed in the  matrix model. Since in the matrix model only mixtures between operators of dimensions differing by an even number can be non-zero (by symmetry of the integral), one has $\langle {\rm Tr}\phi_{\mathbb{S}^4}\rangle=0$, thus recovering the structure in \eqref{mixings}. Finally, once the correct --\textit{i.e.} orthogonalized-- candidates for $\mathcal{O}_{\Delta}^{\mathbb{R}^4}$ have been identified, the different entries of the Zamolodchikov metric  $g_{\Delta_1,\Delta_2}$ are obtained by computing $\langle \mathcal{O}_{\Delta_1}^{\mathbb{R}^4}\,\mathcal{O}_{\Delta_2}^{\mathbb{R}^4}\rangle$ in the matrix model.

A by-product of having identified the correct $\mathcal{O}_{\Delta_1}^{\mathbb{R}^4}$ in terms of  matrix model operators is that one can then also compute correlation functions between CPO's and circular Wilson loops \cite{Rodriguez-Gomez:2016cem}. To that matter one simply evaluates $\langle \mathcal{O}_{\Delta_1}^{\mathbb{R}^4}\,W\rangle$ in the  matrix model using the properly orthogonalized $\mathcal{O}_{\Delta_1}^{\mathbb{R}^4}$.

\end{appendix}

\end{document}